\documentclass[twocolumn]{aastex61}

\newcommand{\diff}{\mathrm{d}}
\newcommand{\Ms}{{\ensuremath{\mathrm{M}_{\odot}}}}
\newcommand{\Rs}{{\ensuremath{\mathrm{R}_{\odot}}}}
\newcommand{\Mpy}{\Ms\,{\rm yr}{\ensuremath{^{-1}}}}
\newcommand{\dm}{\ensuremath{\dot M}}
\newcommand{\gva}{{\sc geneva}}

\newcommand{\Edd}{{\ensuremath{\Gamma_{\rm Edd}}}}
\newcommand{\vsurf}{{\ensuremath{\rm v_{surf}}}}

\newcommand{\vcrit}{{\ensuremath{\rm v_{crit,1}}}}
\newcommand{\vvcrit}{{\ensuremath{\rm v^2_{crit,1}}}}
\newcommand{\vog}{{\ensuremath{\rm v_{crit,2}}}}
\newcommand{\rapv}{{\ensuremath{\vsurf/\vcrit}}}
\newcommand{\rapw}{{\ensuremath{\vsurf/\vog}}}
\newcommand{\Req}{{\ensuremath{R_{\rm eq}}}}
\newcommand{\Rec}{{\ensuremath{R_{\rm eq,crit}}}}
\newcommand{\vvog}{{\ensuremath{\rm v^2_{crit,2}}}}
\newcommand{\RReq}{{\ensuremath{R^2_{\rm eq}}}}
\newcommand{\RRec}{{\ensuremath{R^2_{\rm eq,crit}}}}
\newcommand{\OG}{{\ensuremath{\Omega\Gamma}}-limit}
\newcommand{\Gam}{{\ensuremath{\Gamma_{\Omega}}}}
\newcommand{\prad}{{\ensuremath{P_{\rm rad}}}}
\newcommand\jac{\ensuremath{j_{\rm accr}}}
\newcommand\jk{\ensuremath{j_{\rm K}}}

\received{???}
\revised{???}
\accepted{???}
\submitjournal{ApJL}

\shorttitle{Rotating Supermassive Population III Stars}
\shortauthors{L. Haemmerl\'{e} et al.}

\begin{document}

\title{On the Rotation of Supermassive Stars}

\correspondingauthor{Lionel Haemmerl\'{e}}
\email{lionel.haemmerle@unige.ch}

\author{Lionel Haemmerl\'{e}}
\affiliation{Observatoire de Gen\`eve, Universit\'e de Gen\`eve, chemin des Maillettes 51, CH-1290 Sauverny, Switzerland}

\author{Tyrone E. Woods}
\affiliation{Monash Centre for Astrophysics, School of Physics and Astronomy, Monash University, VIC 3800, Australia}

\author{Ralf S. Klessen}
\affiliation{Universit\"at Heidelberg, Zentrum f\"ur Astronomie, Institut f\"ur Theoretische Astrophysik, Albert-Ueberle-Str. 2, D-69120 Heidelberg, Germany}
\affiliation{Interdisziplin\"ares Zentrum f\"ur wissenschaftliches Rechnen der Universit\"at Heidelberg, Im Neuenheimer Feld 205, D-69120 Heidelberg, Germany}

\author{Alexander Heger}
\affiliation{Monash Centre for Astrophysics, School of Physics and Astronomy, Monash University, VIC 3800,\,Australia}

\author{Daniel J. Whalen}
\affiliation{Institute of Cosmology and Gravitation, University of Portsmouth, Dennis Sciama Building, Portsmouth PO1 3FX, UK}



\begin{abstract}
Supermassive stars born from pristine gas in atomically-cooled haloes are thought to be the progenitors of supermassive black holes at high redshifts.
However, the way they accrete their mass is still an unsolved problem.
In particular, for accretion to proceed, a large amount of angular momentum has to be extracted from the collapsing gas.
Here, we investigate the constraints stellar evolution imposes on this angular momentum problem.
We present an evolution model of a supermassive Population III star including simultaneously accretion and rotation.
We find that, for supermassive stars to form by accretion, the accreted angular momentum has to be about 1\% of the Keplerian angular momentum.
This tight constraint comes from the $\Omega\Gamma$-limit, at which the combination of radiation pressure and centrifugal force cancels gravity.
It implies that supermassive stars are slow rotators, with a surface velocity less than 10 -- 20\% of their first critical velocity,
at which the centrifugal force alone cancels gravity.
At such low velocities, the deformation of the star due to rotation is negligible.
\end{abstract}




\section{Introduction}
\label{sec-in}

Supermassive stars (SMSs), with masses $M>10^4$ \Ms, are candidates for the progenitors of supermassive black holes at high redshifts
(e.g.~\citealt{bromm2003b,begelman2006,regan2009b,agarwal2012,agarwal2017c,latif2013d,latif2013e,regan2014b,dijkstra2014,inayoshi2014a}).
In this scenario, a primordial halo is devoid from H$_2$ molecules due to a strong
external Lyman-Werner radiation field from nearby star-forming regions.
Without this cooling agent, the temperature rises above $\sim10^4$ K, preventing star formation before halo's mass reaches $\sim10^7-10^8$ \Ms.
At this point, the collapse is triggered at rates of $\sim1$ \Mpy, towards a central stellar object \citep{latif2013e,becerra2015,smidt2017}.
Stellar evolution models of Population~III (Pop~III) SMSs show that protostars accreting at these high rates evolve as red supergiants along the Hayashi limit,
with an inflated envelope and a low surface temperature, keeping thus a weak ionising feedback on the accretion flow
\citep{hosokawa2012a,hosokawa2013,haemmerle2017b}.
This allows to maintain accretion onto the protostar towards stellar masses in the supermassive range \citep{hirano2017}.
The star is thought to accrete until $M\simeq3\times10^5$ \Ms, before collapsing into a black hole \citep{umeda2016,woods2017,haemmerle2017b}
due to the general relativistic (GR) instability \citep{chandrasekhar1964}.

Star formation requires processes that remove angular momentum from the inner regions of collapsing pre-stellar clouds,
otherwise the centrifugal force would overcome gravity and prevent further collapse \citep{spitzer1978,bodenheimer1995,maeder2009}.
Numerical studies of Pop~III star formation indicate that primordial protostellar seeds rotate near their critical limit \citep{stacy2011,stacy2013},
at which the centrifugal force exceeds the gravitational attraction.
Above this limit, hydrostatic equilibrium can not be achieved and the star breaks up.
The critical limit is given by:
\begin{equation}
{\vvcrit\over\Rec}={GM\over\RRec}	\quad\Longrightarrow\quad	\vcrit=\sqrt{GM\over\Rec}\,,
\label{eq-vc1}\end{equation}
where \Rec\ is the equatorial radius at the critical limit.
The critical velocity \vcrit, which corresponds to the Keplerian velocity at the stellar surface,
is the maximum rotational velocity a star can reach in hydrostatic equilibrium.
For massive Population~I (Pop~I) stars, the critical limit constrains the accreted angular momentum
to be less than 1/3 of the Keplerian angular momentum \citep{haemmerle2017a}.
If it exceeds this value, internal angular momentum redistribution by convection leads the stellar surface to rotate above the critical limit.

In the present work, we examine similar constraints from stellar evolution on the accretion of angular momentum, in the case of Pop~III SMSs.
These stars evolve at nearly their Eddington limit, at which radiation pressure alone cancels gravity.
However, this limit is not actually reached by SMSs, but approached asymptotically as their masses grow.
Departures from this limit are expressed by the Eddington factor,
\begin{equation}
\Edd:=\left.{\diff\prad\over\diff r}\right/-\rho{GM\over R^2},
\label{eq-edd}\end{equation}
where $\diff\prad/\diff r$ and $\rho$ are the radiation pressure gradient and the mass density at the photosphere,
$G$ is the gravitational constant, and $M$ and $R$ are the stellar mass and radius, respectively.
For a star close to the Eddington limit ($\Edd\gtrsim0.6$), the critical velocity in Eq.~(\ref{eq-vc1}) must be replaced by
\begin{equation}
\vvog=\RReq\cdot2\pi G\bar{\rho}\cdot(1-\Edd),
\label{eq-vc2}\end{equation}
where again \Req\ is the equatorial radius and $\bar{\rho}$ the mean mass density of the star \citep{maeder2000}.
This second critical velocity reflects the contribution from both the centrifugal force and the radiation pressure to counteract gravity.
Similarly, in the presence of rotation, the Eddington factor defined in Eq.~(\ref{eq-edd}) has to be replaced by
\begin{equation}
\Gam=\left.{\diff\prad\over\diff r}\right/\rho\left(-{GM\over R^2}+a_c\right)={\diff\prad\over\diff P}={\Edd\over1-{\Omega^2\over2\pi G\bar{\rho}}},
\label{eq-gam}\end{equation}
where $a_c$ is the centrifugal acceleration, $P$ the total pressure and $\Omega$ the angular velocity.
This is called the \OG\ \citep{maeder2000}.
Eq.~(\ref{eq-vc2}) shows that as $\Edd\rightarrow1$, $\vog\rightarrow0$.

Here, we present a model of an accreting Pop~III SMS, following self-consistently its internal differential rotation,
with the aim of constraining the angular momentum accreted by such objects and establishing their rotational properties.
The stellar evolution code is described in Section~\ref{sec-ge}, and the model in Section~\ref{sec-mod}.
We discuss our results in Section~\ref{sec-dis} and conclude in Section~\ref{sec-out}.

\section{Stellar evolution code}
\label{sec-ge}

The \gva\ code is a one-dimensional hydrostatic stellar evolution code
that numerically solves the four structure equations \citep{eggenberger2008}.
Differential rotation is included, with the assumption of shellular rotation, according to which each isobar rotates as a solid-body \citep{meynet1997}.
Angular momentum transport between the various isobars includes three processes: convection, shear diffusion and meridional circulation.
Convective transport is treated by assuming solid-body rotation in each convective zone.
In radiative regions, the equation of angular momentum transport by shear diffusion and meridional currents reads:
\begin{equation}
\rho\,{\diff(r^2\Omega)\over\diff t}={1\over5r^2}{\diff\over\diff r}\left(\rho r^4\Omega\,U(r)\right)
+{1\over r^2}{\diff\over\diff r}\left(\rho D(r)\,r^4{\diff\Omega\over\diff r}\right),
\label{eq-j}\end{equation}
where $\Omega$ is the angular velocity, $U(r)$ is the amplitude of the radial component of the meridional velocity,
and $D(r)$ is the diffusion coefficient for shear instability (using the prescription of \citealt{maeder1997}).
Accretion is included, at a rate \dm\ fixed externally, as a free parameter.
The thermal properties of the accreted material are those of cold disc accretion \citep{palla1992},
and its rotational properties are fixed externally, through the angular momentum we attribute to the new layers.
In the presence of strong outwards angular momentum transport in the external layers (e.g. a convective envelope),
numerical instabilities prevent full control of the accreted angular momentum.
The code includes also a GR correction to the equation of hydrostatic equilibrium,
the first order post-Newtonian Tolman-Oppenheimer-Volkoff (TOV) correction \citep{fuller1986}.
A general description of the code with rotation, without accretion, can be found in \cite{eggenberger2008}.
The treatment of accretion is detailed in \cite{haemmerle2016a,haemmerle2017a}.

\section{Models}
\label{sec-mod}

\subsection{Inputs to the models}
\label{sec-mod-in}

We consider the fiducial accretion rate
\begin{equation}
\dm=1\ \Mpy,
\end{equation}
typical of SMS formation (e.g.~\citealt{latif2013e,smidt2017}).
We give to the accreted material a fixed fraction of the Keplerian angular momentum,
\begin{equation}
\jac:={\diff J\over\diff M}=f\cdot\jk	\qquad	{\rm with}\ \jk=\sqrt{GMR},
\label{eq-jac}\end{equation}
where $J$ is the angular momentum of the star.
We consider $f=0.01$, i.e. the angular momentum accreted is 1\%\ of the Keplerian angular momentum.

For numerical stability, we start the run at a stellar mass of 10 \Ms.
The initial model is fully convective (polytrope with $n\simeq3/2$, flat entropy profile),
with a radius and central temperature of $R=171$ \Rs\ and $T_c=4\times10^5$ K, respectively.
Because of convection, the initial model rotates as a solid body, whose rotation profile is fully determined by one parameter.
We choose a surface velocity that corresponds to a fraction $f=0.01$ of the critical velocity (Eq.~\ref{eq-vc1}).
The chemical compositions of the initial model and accreted material are identical ($X=0.7516$, $Y=0.2484$ and $Z=0$).
We stop the run at a stellar mass of $10^5$ \Ms, because of numerical instability.

\subsection{Rotational properties of the model}
\label{sec-mod-evol}

\begin{figure}\begin{center}\includegraphics[width=0.49\textwidth]{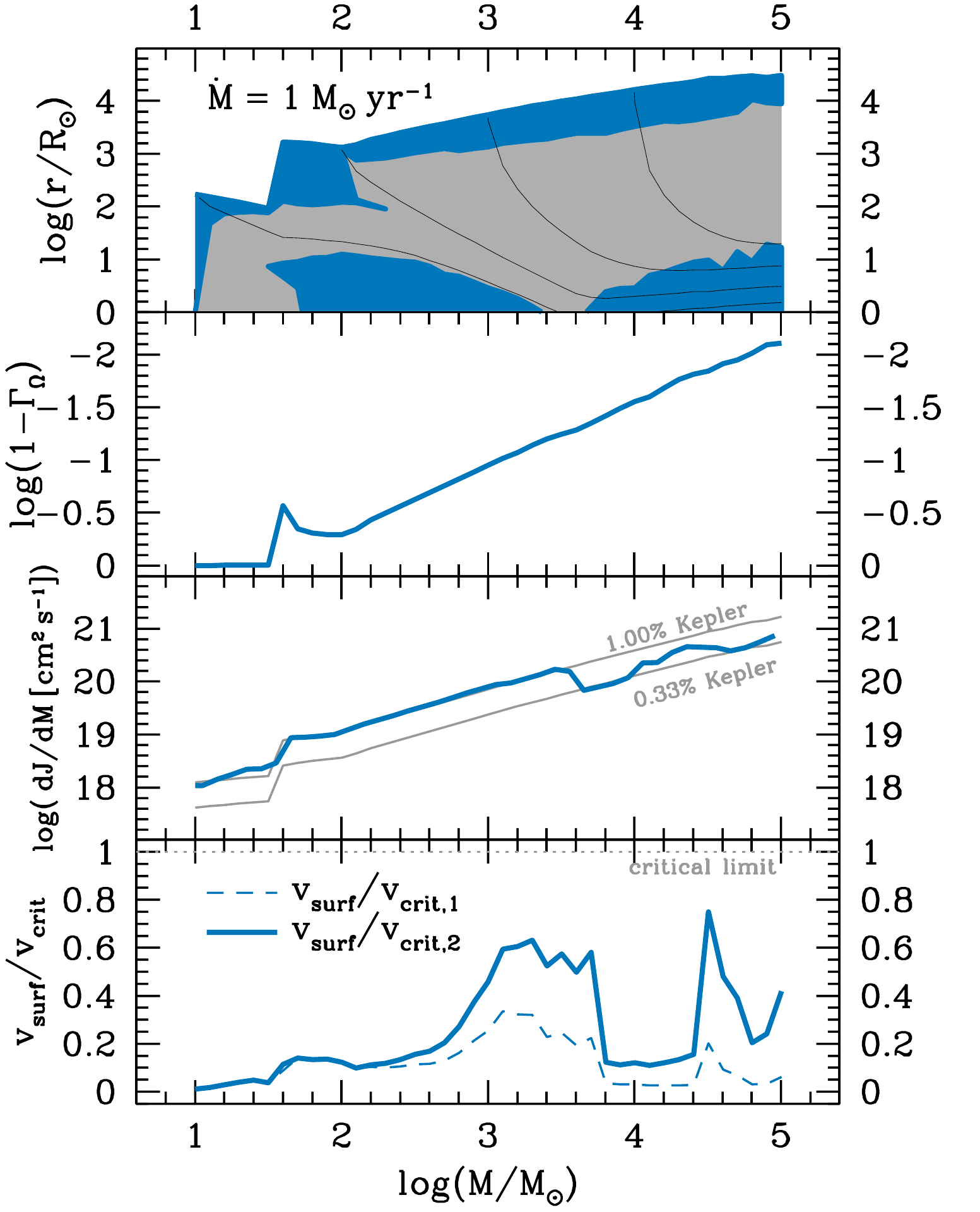}
\caption{Internal structure, Eddington factor, accreted angular momentum and surface velocity
of the model with $\dm=1\,\Mpy$ and $\jac\simeq0.01\jk$, as a function of the current stellar mass.
On the top panel, the upper envelope is the photospheric radius, the coloured regions are convective zones, the grey regions are radiative,
and the black lines indicate the contracting Lagrangian layers of $\log(M_r/\Ms)=1$, 2, 3 and 4.
The Eddington factor shown on the second panel is the corrected Eddington factor in the case with rotation (Eq.~\ref{eq-gam}).
On the third panel, the theoretical \jac\ for $f=1\%$ and 0.33\% is indicated by the grey lines,
while the coloured one is the actual angular momentum accreted by the model.
On the bottom panel, the surface velocity is plotted as a ratio of the critical velocities
\vcrit\ and \vog\ of Eq.~(\ref{eq-vc1}) and (\ref{eq-vc2}).}
\label{fig-rot}\end{center}\end{figure}

Fig.~\ref{fig-rot} shows the evolution of the internal structure, the Eddington factor, the accreted angular momentum
and the surface velocity (ratio to the critical velocities \vcrit\ and \vog) of the model with the inputs of Section~\ref{sec-mod-in}.
The internal structure shown here is identical to that described in \cite{haemmerle2017b}.
The impact of rotation on the internal structure (as well as on the evolutionary track) is negligible.
In the beginning of the evolution, a radiative core forms and grows in mass,
enhancing the internal flux and producing a swelling of the radius by one order of magnitude (luminosity wave, \citealt{larson1972}).
At such high accretion rates, the stellar surface can not contract to the ZAMS, and the radius remains large ($\sim10^4$ \Rs),
growing as $R\propto M^{1/2}$ as the evolution proceeds, in spite of the contraction of each Lagrangian layer.
The star evolves along the Hayashi line, with a convective envelope due to the low temperature ($\sim10^4-10^5$ K) in the inflated regions.
In the centre, after the layers corresponding to the initial model ($M_r<10$ \Ms) have significantly contracted (at $M\simeq3000$ \Ms),
the temperature exceeds $10^8$ K and H-burning becomes efficient, triggering convection.
This convective core grows in mass as the evolution proceeds.

Once the luminosity wave reaches the surface, the Eddington factor exceeds 50\% and eventually evolves as
\begin{equation}
\log(1-\Gam)\simeq1-{2\over3}\,\log{M\over\Ms}.
\label{eq-fitedd}\end{equation}
It reaches 90\% at $M=10^3\,\Ms$ and 99\% at $M=6.3\times10^4\,\Ms$.

Until several 100 \Ms, the rotation velocity remains small ($<20\%$ \vcrit).
But when the star approaches 1000~\Ms, the rapid contraction of the Lagrangian layers
and the instantaneous $J$-transport outwards in the convective envelope make \vsurf\ to increase.
When $\rapw>50\%$, numerical convergence becomes difficult to achieve and we have to decrease \jac\ by a factor of 3.
Several oscillations occur between $\jac=1\%-0.33\%\jk$, leading to oscillations in \rapw\ (10~--~80\%).
The run stops at $M=10^5$\,\Ms\ with $\rapv\simeq10\%$ and $\rapw\simeq40\%$.
Notice that the Eddington factor exceeds 0.6 at $M\simeq100$ \Ms, so that $\rapw>\rapv$ in later stages by a factor of a few.

\begin{figure}\begin{center}\includegraphics[width=0.45\textwidth]{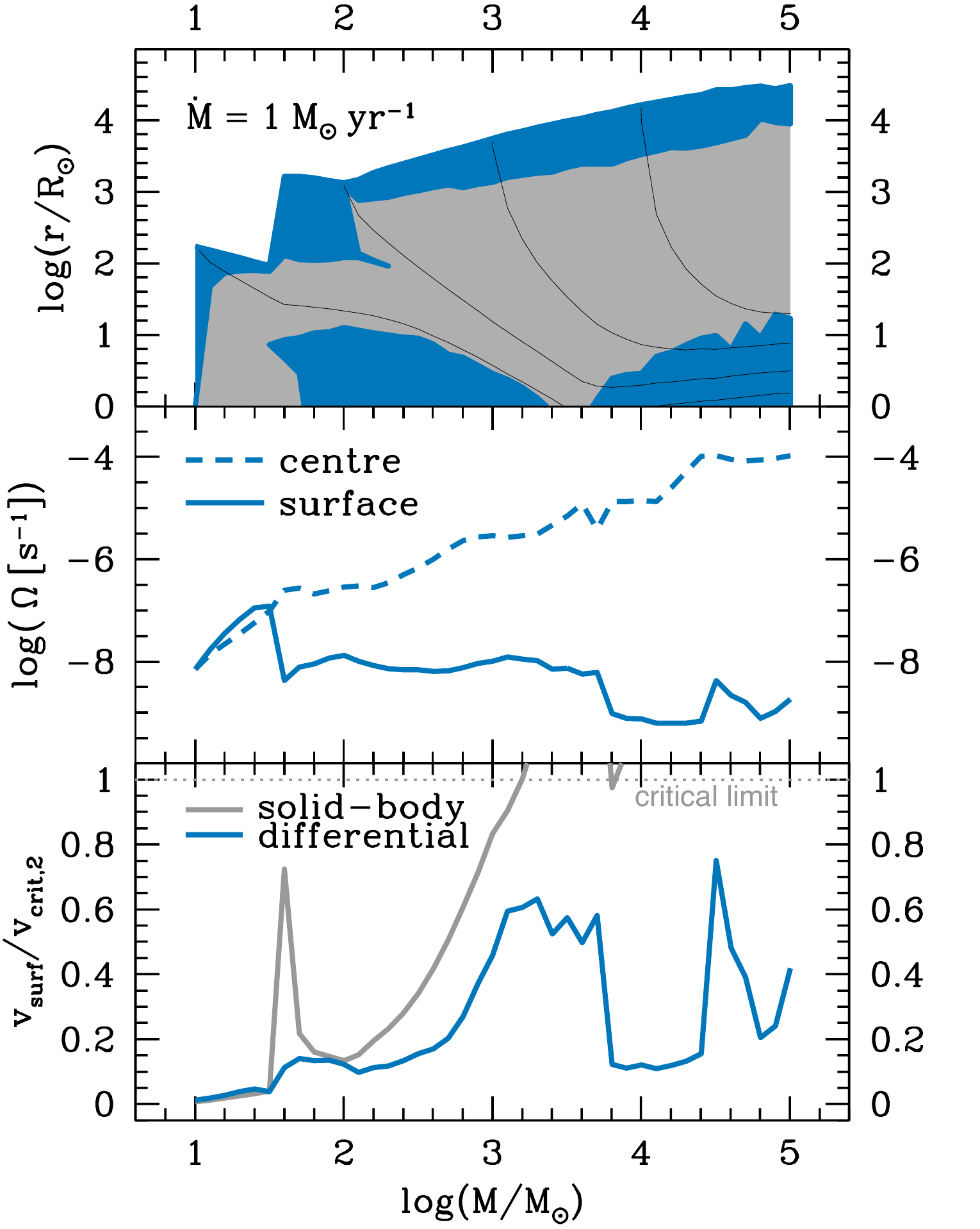}
\caption{Differential rotation in the star, for the same model as Fig.~\ref{fig-rot}.
The upper panel shows again the stellar structure as a point of reference, and is the same as in Fig.~\ref{fig-rot}.
The second panel shows the evolution of the angular velocity in the centre and at the surface.
The bottom panel indicates $\rapw$ of the model, with differential rotation (same as in Fig.~\ref{fig-rot})
and the \rapw\ obtained if we assume solid-body rotation with the same \jac\ 
($\Omega_{\rm sol}:=J/I$, where $I$ is the total moment of inertia of the star).}
\label{fig-dif}\end{center}\end{figure}

In order to see the effect of differential rotation in the model, we plot in Fig.~\ref{fig-dif} the evolution of the angular velocity $\Omega$
in the centre and at the surface as a function of the stellar mass.
We also plot the evolution of \rapw\ of our model, compared to the value obtained with the assumption of solid-body rotation.
In the beginning of the evolution, the $\Omega$ gradient in the growing radiative core is slightly positive outwards.
But differential rotation remains small, and $\Omega$ increases in all the contracting layers, because of local angular momentum conservation.
Once the swelling occurs, the internal contraction departs strongly from homology, and differential rotation develops in the large radiative region.
Indeed, with local angular momentum conservation, one has $\dot\Omega/\Omega=-2\,\dot r/r$,
so that non-homology (i.e. $\dot r/r$ depends on $r$) enhances differential rotation ($\dot\Omega/\Omega$ and thus $\Omega(t)$ depend on $r$).
The angular velocity of the external layers decreases by more than one order of magnitude while that of the centre stops growing temporarily
due to the instantaneous $J$-transport in the transient convective core.
This results in an $\Omega$ gradient that is strongly negative outwards, in particular in the external layers that experience the swelling,
where departures from homology are more pronounced.
When the transient convective core stops growing in mass and contracts in radius with the Lagrangian layers, $\Omega$ increases again in the centre.
In the same time, the surface has converged to the Hayashi limit, and its angular velocity stays nearly constant ($\simeq10^{-8}$~s$^{-1}$).
Thus, differential rotation is enhanced by the central contraction and $\Omega$ soon differs by 3 -- 4 orders of magnitude between the centre and the surface.
The spinning-up of the centre stops only when H-burning becomes efficient enough to halt contraction in the convective core, at 20\,000 -- 30\,000 \Ms.
Then the ratio of $\Omega$ in the core and at the surface stabilises at a level of 4 -- 5 orders of magnitude, depending on the variations in \jac.

Angular momentum transport by shear diffusion remains negligible during the whole evolution because of the short evolutionary timescale.
This is also true for meridional circulation, except in a small region under the convective envelope.
During the entire evolution, the internal angular momentum distribution is dominated by convective transport and local angular momentum conservation.
Since large regions of the star are radiative and contraction is far from homology, differential rotation is extremely strong.

The impact of differential rotation on the evolution of the surface velocity is shown in the bottom panel of Fig.~\ref{fig-dif}.
While for differential rotation \rapw\ remains in the range 10~--~80\%,
the value of \rapw\ computed by assuming rigid rotation reaches the critical limit before the star becomes supermassive.
After a short peak at 70\% during the swelling, this ratio falls to 10\%
and then grows rapidly due to the increase in \Edd\ (through the decrease in \vog, Eq.~\ref{eq-vc2}).
The \OG\ is reached at 1500~\Ms, that is long before the star enters the supermassive regime.

\section{Discussion}
\label{sec-dis}

\subsection{Semi-analytical interpretation}
\label{sec-dis-fit}

\begin{figure}\begin{center}\includegraphics[width=0.45\textwidth]{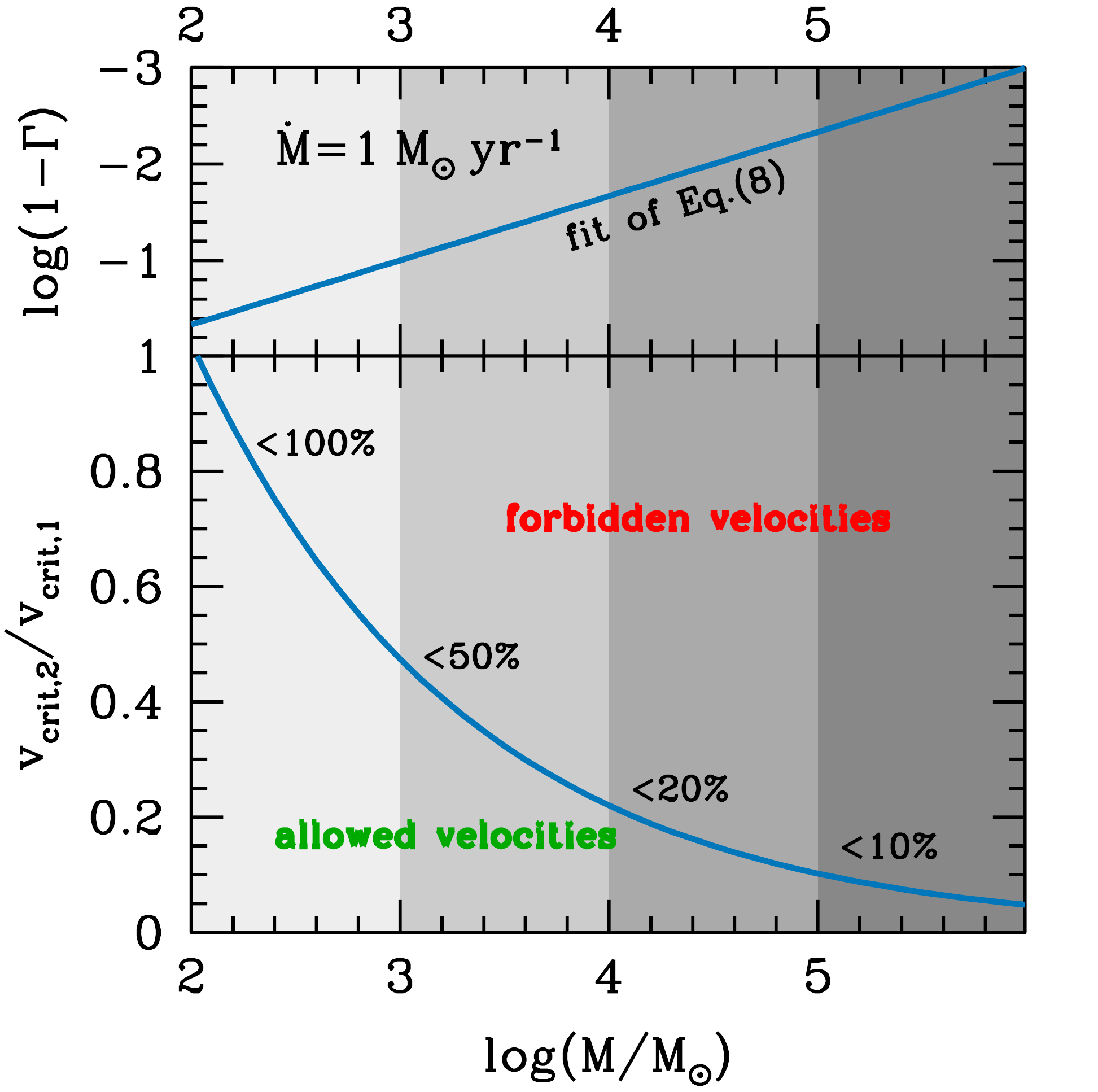}
\caption{Semi-analytical interpretation of the constraint from the \OG.
The upper panel shows the fit of Eq.~(\ref{eq-fitedd}).
The lower panel shows the resulting \vog/\vcrit\ vs. $M$ relation (Eq.~\ref{eq-fit})
that delimitates the allowed and forbidden surface rotation velocities.}
\label{fig-fit}\end{center}\end{figure}

If we neglect the deformation of the star ($\Req=R$), which is justified for $\rapv\lesssim50\%$,
the corrected critical velocity from Eq.~(\ref{eq-vc2}) can be written as
\begin{equation}
\vog={3\over2}\vcrit\ \sqrt{1-\Edd},
\label{eq-vc3}\end{equation}
since $\bar{\rho}=M_r/{4\over3}\pi R^3$.
The model described in Section~\ref{sec-mod-evol} shows that the $\Edd-M$ relation can be fitted by a power-law (Eq.~\ref{eq-fitedd}).
Using this fit in Eq.~(\ref{eq-vc3}), one obtains
\begin{equation}
{\vog\over\vcrit}\simeq4.7\times\left({M\over\Ms}\right)^{-1/3},
\label{eq-fit}\end{equation}
which gives $\vog/\vcrit=22\%$ for $M=10^4$ \Ms, and $\vog/\vcrit=10\%$ for $M=10^5$ \Ms.
Thus, a SMS has to rotate at less than 10 -- 20\% of its classical critical velocity \vcrit, otherwise it would exceed \vog\ and would reach the \OG.
This is in agreement with the model in Fig.~\ref{fig-rot}.
We illustrate this semi-analytical interpretation in Fig.~\ref{fig-fit}.

\subsection{Constraints on the accretion of angular momentum}
\label{sec-dis-j}

The model described above shows that for accreting SMSs to avoid the \OG\
their accreted angular momentum must be of the order of 1\% of the Keplerian angular momentum.
It indicates that SMS formation by accretion requires strong mechanisms for extracting angular momentum from the accretion disc
like magnetic fields \citep{schober2012,latif2016a}, viscosity \citep{popham1991,takahashi2017} or gravitational torques \citep{wise2008a}.
This is a particular case of the classical angular momentum problem \citep{spitzer1978,bodenheimer1995,maeder2009}.
The constraint obtained here is much stronger than in the case of massive Pop~I stars ($\jac\lesssim1/3$~\jk, \citealt{haemmerle2017a}).

On the other hand, the effect of differential rotation softens the constraint on \jac.
Assuming solid-body rotation, \cite{lee2016} computed the rotational properties of Pop III stars accreting at $\dm\sim10^{-3}$ \Mpy.
They found that, even for $\jac=0.01\jk$, the \OG\ is reached at $M=20-40$ \Ms, preventing more massive stars from forming by accretion.
Due to the differences in the initial conditions and the accretion rates, a direct comparison of the early evolution is not possible.
However, our model shows that solid-body rotation requires $\jac\ll0.01\jk$ for the star to become supermassive by accretion,
while with differential rotation $\jac=0.01\jk$ is low enough.
This suggests that the constraint on \jac\ obtained by \cite{lee2016} is artificially strong due to their assumption of solid-body rotation.

What happens once the star reaches the \OG?
According to the stationary disc models of \cite{takahashi2017}, viscosity is able to maintain the advection of angular momentum arbitrarily low.
Their models show that stationary solutions exist for any $M$ and \dm\ with a \jac\ as close to 0 as required,
provided that the surface density in the inner disc is high enough.
Thus, if at a given point the accretion process is stopped by the \OG,
the mass that rotates too fast will accumulate in the disc until viscosity is efficient enough to extract the excess of angular momentum.
If this picture is correct, the star-disc system will adjust to a configuration corresponding to the highest \jac\ compatible with the \OG.
Our model can then be seen as a lower limit for \jac\ and \vsurf, because \rapw\ remains always lower than 80\%.

\section{Conclusion}
\label{sec-out}

We have described the first stellar evolution model of Pop~III SMS that simultaneously include accretion and rotation.
We obtained the following results:

\begin{enumerate}

\item{\it Supermassive stars have to be slow rotators.}
Since SMSs evolve at close to the Eddington limit, the \OG\ imposes tight constraints on their rotation velocity.
For $M>10^4$\,\Ms, the rotation velocity can not exceed $\sim20\%$ of the first critical velocity (Eq.~\ref{eq-vc1}).
For $M>10^5$\,\Ms, the limit is $\sim10\%$.
For such slow rotators, the impact of rotation on the stellar structure is negligible.

\item{\it Supermassive star formation by accretion requires mechanisms efficient enough to remove most ($\sim99\%$) of the angular momentum from the accretion disc.}
Indeed, the constraint on the rotation velocity translates into a constraint on the accreted angular momentum,
which must not exceed 1\% of the Keplerian value.
The main mechanisms expected to play this role are viscosity, magnetic fields and gravitational torques from spiral arms in the accretion disc.
If the angular momentum accreted remains slightly lower than $1\%$ of the Keplerian angular momentum,
then accretion can proceed towards $M>10^5$\,\Ms\ without facing the \OG.

\item{\it Supermassive stars forming by accretion rotate highly differentially, with a frequency in the core 4 or 5 orders of magnitude higher than in the envelope.}
This is due to the highly non-homologous nature of internal stellar contraction.
In radiative regions, the rotation profile is dominated by local angular momentum conservation,
so that departures from homology result in departures from solid-body rotation.
Differential rotation softens the constraint on the accreted angular momentum
because it allows more angular momentum to be contained in the central regions while keeping a slow rotation velocity at the surface.

\end{enumerate}

\section*{Acknowledgements}

LH and RSK were supported by the European Research Council under the European Community's Seventh Framework Programme (FP7/2007 - 2013)
via the ERC Advanced Grant `STARLIGHT: Formation of the First Stars' (project number 339177).
Part of this work was supported by the Swiss National Science Foundation.
DJW was supported by STFC New Applicant Grant ST/P000509/1.


\begin{thebibliography}{}
\expandafter\ifx\csname natexlab\endcsname\relax\def\natexlab#1{#1}\fi
\providecommand{\url}[1]{\href{#1}{#1}}

\bibitem[{{Agarwal} {et~al.}(2012){Agarwal}, {Khochfar}, {Johnson}, {Neistein},
  {Dalla Vecchia}, \& {Livio}}]{agarwal2012}
{Agarwal}, B., {Khochfar}, S., {Johnson}, J.~L., {et~al.} 2012, \mnras, 425,
  2854

\bibitem[{{Agarwal} {et~al.}(2017){Agarwal}, {Regan}, {Klessen}, {Downes}, \&
  {Zackrisson}}]{agarwal2017c}
{Agarwal}, B., {Regan}, J., {Klessen}, R.~S., {Downes}, T.~P., \& {Zackrisson},
  E. 2017, \mnras, 470, 4034

\bibitem[{{Becerra} {et~al.}(2015){Becerra}, {Greif}, {Springel}, \&
  {Hernquist}}]{becerra2015}
{Becerra}, F., {Greif}, T.~H., {Springel}, V., \& {Hernquist}, L.~E. 2015,
  \mnras, 446, 2380

\bibitem[{{Begelman} {et~al.}(2006){Begelman}, {Volonteri}, \&
  {Rees}}]{begelman2006}
{Begelman}, M.~C., {Volonteri}, M., \& {Rees}, M.~J. 2006, \mnras, 370, 289

\bibitem[{{Bodenheimer}(1995)}]{bodenheimer1995}
{Bodenheimer}, P. 1995, \araa, 33, 199

\bibitem[{{Bromm} \& {Loeb}(2003)}]{bromm2003b}
{Bromm}, V., \& {Loeb}, A. 2003, \apj, 596, 34

\bibitem[{{Chandrasekhar}(1964)}]{chandrasekhar1964}
{Chandrasekhar}, S. 1964, \apj, 140, 417

\bibitem[{{Dijkstra} {et~al.}(2014){Dijkstra}, {Ferrara}, \&
  {Mesinger}}]{dijkstra2014}
{Dijkstra}, M., {Ferrara}, A., \& {Mesinger}, A. 2014, \mnras, 442, 2036

\bibitem[{{Eggenberger} {et~al.}(2008){Eggenberger}, {Meynet}, {Maeder},
  {Hirschi}, {Charbonnel}, {Talon}, \& {Ekstr{\"o}m}}]{eggenberger2008}
{Eggenberger}, P., {Meynet}, G., {Maeder}, A., {et~al.} 2008, \apss, 316, 43

\bibitem[{{Fuller} {et~al.}(1986){Fuller}, {Woosley}, \& {Weaver}}]{fuller1986}
{Fuller}, G.~M., {Woosley}, S.~E., \& {Weaver}, T.~A. 1986, \apj, 307, 675

\bibitem[{{Haemmerl{\'e}} {et~al.}(2016){Haemmerl{\'e}}, {Eggenberger},
  {Meynet}, {Maeder}, \& {Charbonnel}}]{haemmerle2016a}
{Haemmerl{\'e}}, L., {Eggenberger}, P., {Meynet}, G., {Maeder}, A., \&
  {Charbonnel}, C. 2016, \aap, 585, A65

\bibitem[{{Haemmerl{\'e}} {et~al.}(2017{\natexlab{a}}){Haemmerl{\'e}},
  {Eggenberger}, {Meynet}, {Maeder}, {Charbonnel}, \&
  {Klessen}}]{haemmerle2017a}
{Haemmerl{\'e}}, L., {Eggenberger}, P., {Meynet}, G., {et~al.}
  2017{\natexlab{a}}, \aap, 602, A17

\bibitem[{{Haemmerl{\'e}} {et~al.}(2017{\natexlab{b}}){Haemmerl{\'e}}, {Woods},
  {Klessen}, {Heger}, \& {Whalen}}]{haemmerle2017b}
{Haemmerl{\'e}}, L., {Woods}, T.~E., {Klessen}, R.~S., {Heger}, A., \&
  {Whalen}, D.~J. 2017{\natexlab{b}}, ArXiv e-prints, arXiv:1705.09301

\bibitem[{{Hirano} {et~al.}(2017){Hirano}, {Hosokawa}, {Yoshida}, \&
  {Kuiper}}]{hirano2017}
{Hirano}, S., {Hosokawa}, T., {Yoshida}, N., \& {Kuiper}, R. 2017, Science,
  357, 1375

\bibitem[{{Hosokawa} {et~al.}(2012){Hosokawa}, {Omukai}, \&
  {Yorke}}]{hosokawa2012a}
{Hosokawa}, T., {Omukai}, K., \& {Yorke}, H.~W. 2012, \apj, 756, 93

\bibitem[{{Hosokawa} {et~al.}(2013){Hosokawa}, {Yorke}, {Inayoshi}, {Omukai},
  \& {Yoshida}}]{hosokawa2013}
{Hosokawa}, T., {Yorke}, H.~W., {Inayoshi}, K., {Omukai}, K., \& {Yoshida}, N.
  2013, \apj, 778, 178

\bibitem[{{Inayoshi} {et~al.}(2014){Inayoshi}, {Omukai}, \&
  {Tasker}}]{inayoshi2014a}
{Inayoshi}, K., {Omukai}, K., \& {Tasker}, E. 2014, \mnras, 445, L109

\bibitem[{{Larson}(1972)}]{larson1972}
{Larson}, R.~B. 1972, \mnras, 157, 121

\bibitem[{{Latif} \& {Schleicher}(2016)}]{latif2016a}
{Latif}, M.~A., \& {Schleicher}, D.~R.~G. 2016, \aap, 585, A151

\bibitem[{{Latif} {et~al.}(2013{\natexlab{a}}){Latif}, {Schleicher}, {Schmidt},
  \& {Niemeyer}}]{latif2013d}
{Latif}, M.~A., {Schleicher}, D.~R.~G., {Schmidt}, W., \& {Niemeyer}, J.
  2013{\natexlab{a}}, \mnras, 433, 1607

\bibitem[{{Latif} {et~al.}(2013{\natexlab{b}}){Latif}, {Schleicher}, {Schmidt},
  \& {Niemeyer}}]{latif2013e}
{Latif}, M.~A., {Schleicher}, D.~R.~G., {Schmidt}, W., \& {Niemeyer}, J.~C.
  2013{\natexlab{b}}, \mnras, 436, 2989

\bibitem[{{Lee} \& {Yoon}(2016)}]{lee2016}
{Lee}, H., \& {Yoon}, S.-C. 2016, \apj, 820, 135

\bibitem[{{Maeder}(1997)}]{maeder1997}
{Maeder}, A. 1997, \aap, 321, 134

\bibitem[{{Maeder}(2009)}]{maeder2009}
---. 2009, {Physics, Formation and Evolution of Rotating Stars} (Springer),
  doi:10.1007/978-3-540-76949-1

\bibitem[{{Maeder} \& {Meynet}(2000)}]{maeder2000}
{Maeder}, A., \& {Meynet}, G. 2000, \aap, 361, 159

\bibitem[{{Meynet} \& {Maeder}(1997)}]{meynet1997}
{Meynet}, G., \& {Maeder}, A. 1997, \aap, 321, 465

\bibitem[{{Palla} \& {Stahler}(1992)}]{palla1992}
{Palla}, F., \& {Stahler}, S.~W. 1992, \apj, 392, 667

\bibitem[{{Popham} \& {Narayan}(1991)}]{popham1991}
{Popham}, R., \& {Narayan}, R. 1991, \apj, 370, 604

\bibitem[{{Regan} \& {Haehnelt}(2009)}]{regan2009b}
{Regan}, J.~A., \& {Haehnelt}, M.~G. 2009, \mnras, 396, 343

\bibitem[{{Regan} {et~al.}(2014){Regan}, {Johansson}, \& {Wise}}]{regan2014b}
{Regan}, J.~A., {Johansson}, P.~H., \& {Wise}, J.~H. 2014, \apj, 795, 137

\bibitem[{{Schober} {et~al.}(2012){Schober}, {Schleicher}, {Federrath},
  {Glover}, {Klessen}, \& {Banerjee}}]{schober2012}
{Schober}, J., {Schleicher}, D., {Federrath}, C., {et~al.} 2012, \apj, 754, 99

\bibitem[{{Smidt} {et~al.}(2017){Smidt}, {Whalen}, {Johnson}, \&
  {Li}}]{smidt2017}
{Smidt}, J., {Whalen}, D.~J., {Johnson}, J.~L., \& {Li}, H. 2017,
  arXiv:1703.00449, arXiv:1703.00449

\bibitem[{{Spitzer}(1978)}]{spitzer1978}
{Spitzer}, L. 1978, {Physical processes in the interstellar medium} (New York:
  Wiley-Inter science), doi:10.1002/9783527617722

\bibitem[{{Stacy} {et~al.}(2011){Stacy}, {Bromm}, \& {Loeb}}]{stacy2011}
{Stacy}, A., {Bromm}, V., \& {Loeb}, A. 2011, \mnras, 413, 543

\bibitem[{{Stacy} {et~al.}(2013){Stacy}, {Greif}, {Klessen}, {Bromm}, \&
  {Loeb}}]{stacy2013}
{Stacy}, A., {Greif}, T.~H., {Klessen}, R.~S., {Bromm}, V., \& {Loeb}, A. 2013,
  \mnras, 431, 1470

\bibitem[{{Takahashi} \& {Omukai}(2017)}]{takahashi2017}
{Takahashi}, S.~Z., \& {Omukai}, K. 2017, \mnras, 472, 532

\bibitem[{{Umeda} {et~al.}(2016){Umeda}, {Hosokawa}, {Omukai}, \&
  {Yoshida}}]{umeda2016}
{Umeda}, H., {Hosokawa}, T., {Omukai}, K., \& {Yoshida}, N. 2016, \apjl, 830,
  L34

\bibitem[{{Wise} {et~al.}(2008){Wise}, {Turk}, \& {Abel}}]{wise2008a}
{Wise}, J.~H., {Turk}, M.~J., \& {Abel}, T. 2008, \apj, 682, 745

\bibitem[{{Woods} {et~al.}(2017){Woods}, {Heger}, {Whalen}, {Haemmerl{\'e}}, \&
  {Klessen}}]{woods2017}
{Woods}, T.~E., {Heger}, A., {Whalen}, D.~J., {Haemmerl{\'e}}, L., \&
  {Klessen}, R.~S. 2017, \apjl, 842, L6

\end{thebibliography}
\end{document}